# BIG DATA ANALYTICS IN HUMANITARIAN AND DISASTER OPERATIONS: A SYSTEMATIC REVIEW

Abhilash Kondraganti, University of Liverpool, abhilash.kondraganti@liverpool.ac.uk

**Abstract:** By the outset of this review, 168 million people needed humanitarian aid, and the number grew to 235 million by the end of the completion of this review. There is no time to lose, definitely no data to lose. Humanitarian relief is crucial not just to contend with a pandemic once a century but also to provide help during civil conflicts, ever-increasing natural disasters, and other forms of crisis. Reliance on technology has never been so relevant and critical than now. The creation of more data and advancements in data analytics provides an opportunity to the humanitarian field. This review aimed at providing a holistic understanding of big data analytics in a humanitarian and disaster setting. A systematic literature review method is used to examine the field and the results of this review explain research gaps, and opportunities available for future research. This study has shown a significant research imbalance in the disaster phase, highlighting how the emphasis is on responsive measures than preventive measures. Such reactionary measures would only exacerbate the disaster, as is the case in many nations with COVID-19. Overall this research details the current state of big data analytics in a humanitarian and disaster setting.

## 1. INTRODUCTION

The humanitarian crisis has become one of the most urgent issues to be addressed in recent times, especially in 2020. One humanitarian crisis can undo years of growth, with economic consequences exacerbated by the rise in unemployment and implications such as poverty. Human suffering is ever-increasing and once in a century crisis like COVID-19 will make it more difficult for humanitarian organizations to reach the affected people. According to the United Nations Office for the Coordination of Humanitarian Affairs (UN OCHA), we are looking at 235 million people across the globe who need humanitarian assistance in 2021, a sharp increase of 40% compared to last year (UN OCHA, 2020a). Human suffering at large is caused by either natural or human-induced disasters, but these overwhelming numbers are usually due to natural disasters. There were 7,348 natural disasters in the period 2000-2019 which is more than one reported natural disaster per day over the last twenty years, while this number for the period between 1980-1999 was just above 0.5 disasters per day (UNDRR, 2020).

Humanitarian actors such as local and international NGOs are working closely with governments and United Nation agencies such as OCHA to reach people affected by these disasters. The pooling of funds for relief efforts is always going to be an issue as it is uncertain, but more importantly, it is a challenge to use the raised funds efficiently. In 2019, OCHA required $21.9 billion to alleviate the suffering (UN OCHA, 2019) and it has successfully received $18 billion as contributions from the donors (UN OCHA, 2020b). This suggests that the funding deficit is declining relative to previous years, however, considering the unpredictable nature of donations, humanitarian actors must concentrate on cost-effectiveness in operations to prevent excessive reliance solely on funding. Also, disaster relief is mainly made up of logistical operations, as per Van Wassenhove (2006), it is about 80%. After 15 years of this assertion, one of the findings of an applied research project reaffirmed that the representation of humanitarian logistics is in the range of between 60% - 80% of overall expenditure on humanitarian relief (Lacourt and Radosta, 2019). Further, 35% - 40% of these logistical funds are wasted due to a lack of analysis on spend and duplication of efforts (Day et al., 2012; Kwapong Baffoe and Luo, 2020). Being unable to raise funds may be attributable to economic or diplomatic causes, but poor utilisation of the funds is down to logistical and operational failure needing serious attention.

A key approach to resolve efficiency and effectiveness issues in logistical operations has been the application of new technologies. Rapid advances in technology and the use of Big Data Analytics (BDA) has created new opportunities for many industries (McKinsey & Company, 2018). Swaminathan's (2018) opinion is that traditional humanitarian organisations may also experience improvements with the integration of BDA and are better placed along with profit-driven corporations to make use of this technology. On the other hand, Sharma and Joshi (2019) argue that data does not reflect the actual situation of the field, and over-reliance on big data technologies could impact humanitarian operations. It might also cost one of the basic humanitarian principles, being humane (UN OCHA, 2012) which might be difficult to accomplish in a non-human data-driven engagement. Experts involved in disaster response however seem to favour the use of data analytics or a system that may be less challenging for them in areas such as analysing the credibility of data, or enhancement in identifying the disaster location (Thom et al., 2016). This is justified due to the critical, inherently unpredictable, and complex nature of operations on the field needing quick decision making (Knox Clarke and Campbell, 2020). More importantly, the humanitarian and disaster operations (HDO) field is becoming more diversified with the inclusion of individuals such as volunteers, people working in crowd sourcing who are outside of humanitarian organizations and are not equipped with enough training.

Review of the existing literature shows a considerable gap between the business and humanitarian sectors in attention to and use of BDA. While the business sector has made significant advances in shifting from descriptive analytics to predictive analytics, and more recently introduction of sophisticated prescriptive analytics (Lepenioti et al., 2020) (e.g. UPS (Delen, 2019) and PopSugar (van Rijmenam, 2019)), such applications in humanitarian and disaster sector has been rare (Centre for Humanitarian Data, 2019).

The disparity in the academic research towards BDA in HDO needs to be thoroughly reviewed. Therefore, this study aims to investigate the role of BDA in HDO, deliver a holistic understanding, and see where the research can be strengthened. The objective of this research is to review and analyse how BDA has been used in various disasters, disaster phases, and categories in the humanitarian field. The three research questions for this study are stated below:

RQ1:　How has the research on the application of BDA for HDO evolved over time?

RQ2:　What is the status of the BDA application across different disaster categories, disaster phases, disaster locations, and what different types of big data have been used?

RQ3:　What are the key theoretical lenses applied to examine BDA application in HDO?

First, this review provides the methodology adopted to conduct the research, which is explained in the next section with the help of the review protocol, search strategy, and assessing the quality of articles. In the following sections, the results are presented and discussed the key aspects of the review. Finally, this review provides the directions for future research to advance the use of BDA in the HDO field, and reflect on the limitations of the review.

## 2.　METHODOLOGY

A systematic literature review (SLR) is employed as a research method to collect and critically assess the existing knowledge in the research field to respond to the research questions. The selection of SLR is based on four compelling reasons; Firstly, it aims to add clarity to the entire process with the support of review protocol, and a strategically designed search strategy (Booth et al., 2012).

Secondly, the researcher aspires to avoid any bias including selection and publication bias in conducting the review and the principles of SLR can minimize this and facilitates in generating more reliable outcomes (Becheikh et al., 2006). Thirdly, the ability to be transparent in the whole review process (Booth et al., 2012) and finally, it could be reproducible for other scholars interested in further exploring this research (Booth et al., 2012). This review has broadly followed the guidelines of Tranfield et al. (2003) and Denyer and Tranfield (2009), two widely used SLR methods in the management discipline.

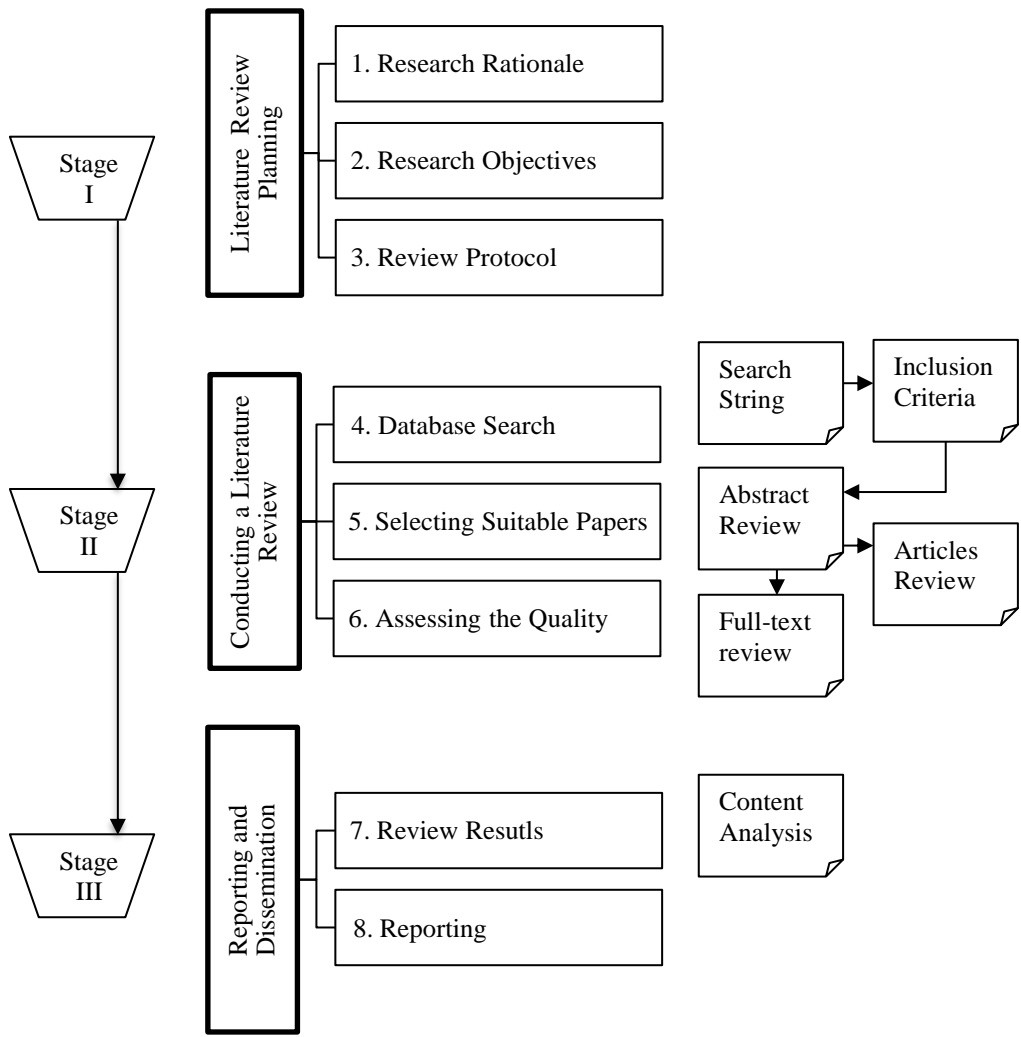

**Figure 1 Systematic literature review process**

## 2.1. Review Protocol

The research protocol helps in performing the second stage of the review 'conducting a literature review' which is the core part of this study in the SLR process shown in figure 1. The purpose of this protocol is to avoid any researcher bias (Tranfield et al., 2003), hence the search strategy is in place with a formal set of rules put in place to identify the suitable articles for this study. So, the quest for extant literature is enabled by the selection of a more fitting citation database, and Scopus is chosen for this study. Scopus is considered the largest multidisciplinary database and has more coverage of journals than Web of Science (Aghaei Chadegani et al., 2013). Also, the search results are integrated from other databases including Springer Link, Science Direct, Wiley Online Library, Emerald Insight, etc. (Roy et al., 2018).

## 2.2. Search Strategy

The efficacy of SLR is dependent on the search strategy that is used to shortlist the scholarly papers by implementing inclusion and exclusion criteria (Snyder, 2019). In Scopus, a search string was generated with the aid of the Boolean operators, which reflects both BDA and HDO in the search results. Here, the researcher is mindful of using more keywords as it could narrow down the search significantly and possibly omit any relevant documents; therefore, the search string is not tight and left as generic as possible. This is a nascent field, and as a measure, the author is very cautious in selecting the search keywords. BDA is divided into two keywords: 'big data' and 'analytics', since some articles may use either name in the title, abstract, or keywords instead of big data analytics. Furthermore, these keywords are paired with another set of keywords "humanitarian" and "disaster" to encompass the entire HDO field. The exact search string used is given below:

(("analytics" AND "humanitarian") OR ("analytics" AND "disaster") OR ("big data" AND "humanitarian") OR ("big data" AND "disaster"))

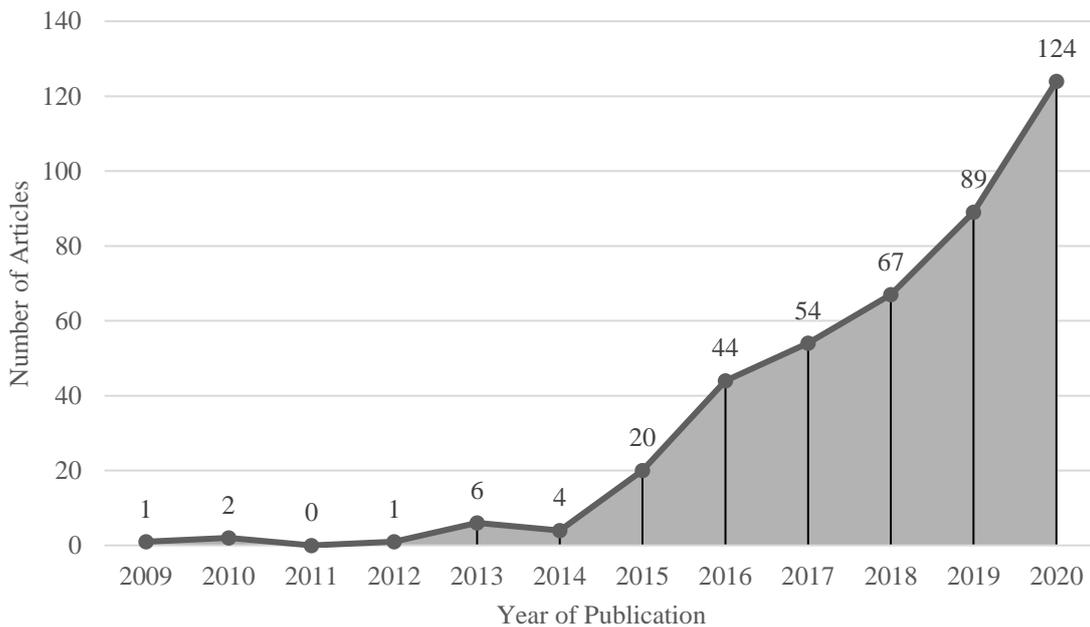

**Figure 2 Research capacity over the years by the number of articles. Source: Compilation by Author**

The search strategy as shown in table 1, comprises 5 levels that have assisted in selecting the suitable articles and these are carried out within the Scopus. The use of the search string in the search area culminated in 1,563 documents in the first level. As the study area is multidisciplinary, 5 relevant subject areas are added at the second level and the number is down to 1,354. This study considers only peer-reviewed articles, which applied at level three significantly reduces the number to 483. The reasoning behind considering only the published work is it can improve the quality of the review since most publications adopt a meticulous peer-review process (Light and Pillemer, 1984). Furthermore, only journal articles are retained in level four but there is minimal change in the number of articles which has gone down to 468. In the end, filtering to articles in the English language leaves us with the final list of 417 articles. There was no restriction on publication year in the search, however, the oldest published article is traced back to the year 2009 as shown in figure 2. The data collection was started in April 2020, later the results were updated in July and in December same year. Scopus search results for this review are up to date as of 31st December 2020.

| Level | Criteria | Description | Results |
|---|---|---|---|
| L1 | Search area | Title, Key Word, Abstract | n = 1,563 |
| L2 | Subject area | Computer Science, Engineering, Decision Sciences, Social Sciences, Business, Management and Accounting | n = 1,354 |
| L3 | Document type | Article, Review | n = 483 |
| L4 | Source type | Journal | n = 468 |
| L5 | Language | English | n = 417 |

**Table 1 Search criteria results in Scopus. Source: Compilation by Author**

## 2.3. Abstract and Full-Text Review

A further shortlisting procedure is applied by reviewing the results from the search (inclusion and exclusion) criteria. The abstracts for 417 articles are reviewed thoroughly, however when the author felt that the abstract information of the article is not sufficient enough to determine the significance, the full-text review was performed. This process omitted approximately 61% of the articles and 160 articles are retained for the full-text review. This review is concept-centric in which a structure is developed to note down the key concepts for each article to obtain comprehensiveness from the associated literature (Webster and Watson, 2002). The inclusion criterion of full-text articles is solely assessed based on one single criterion; 'Is the article at the intersection of BDA and HDO?' This assessment is carried out by dividing each article into three different categories. As shown in table 2, category 1 represents the most relevant articles to the research area. Wherein category 2 is relevant to some extent in bringing BDA and HDO together. And the final category 3 articles are not related and do not contribute to the progress of this research. This review considered categories 1 and 2 articles which are 86.

| Category | Description | Results |
|---|---|---|
| 1 | The focus of the article is on HDO and BDA as a key point. | n = 54 |
| 2 | Considerable insights in the article on the intersection of BDA with HDO. | n = 32 |
| 3 | The article is not relevant to the research area. | n = 74 |

**Table 2 Full-text review results. Source: Compilation by Author**

## 3. RESULTS

In this section, the author presents key findings from the final list of articles meticulously selected after quality assessment and these are presented chronologically. While there are 86 final papers in total, the author chose to exclude 13 review papers to obtain clear outcomes. As a result, 73 articles have been evaluated.

## 3.1. Disaster Categories

The disaster classification shown in figure 3 illustrates that scholars place a greater emphasis on natural disasters, as natural occurrences account for more than half of disasters in the reviewed papers. Within the natural disaster generic group, geophysical disasters, including earthquakes and tsunamis, comprise 15% and hydrological disasters, including floods and heavy rains, account for 14% of the total. In this, both floods and earthquakes are the predominant choices for researchers. The meteorological group of hurricanes and typhoons with 12% is another disaster group that has been prioritized to resolving the challenges. However, the climatological disaster group is relatively less explored with 6%, and biological group research is insignificant with 1%. A few scholars focused on multiple disaster groups within natural disasters in their work with 3% coverage. The remaining papers in the natural disaster group are generic and not aimed at any particular disaster group, accounting for 4%. This takes the overall interest in natural disasters by academics to 55%.

Though, the human-induced disaster generic group is less discussed by researchers, with just 8%. According to the Swiss Re's (2019) report, 37% of disasters reported in 2018 are human-induced and the 10-year average is more than 30% but the scholars' focus in this area is trivial. Several articles did not cover either of the disaster groups with a significant 22% coverage, and this contains conceptual papers and empirical papers mainly related to generic humanitarian supply chain, ethics, and privacy. The technology was tested in a non-disaster environment in one paper, so it was not assigned to any of the disasters. Figure 5 shows the remaining 14% of studies on the combination of both natural and human-induced disasters.

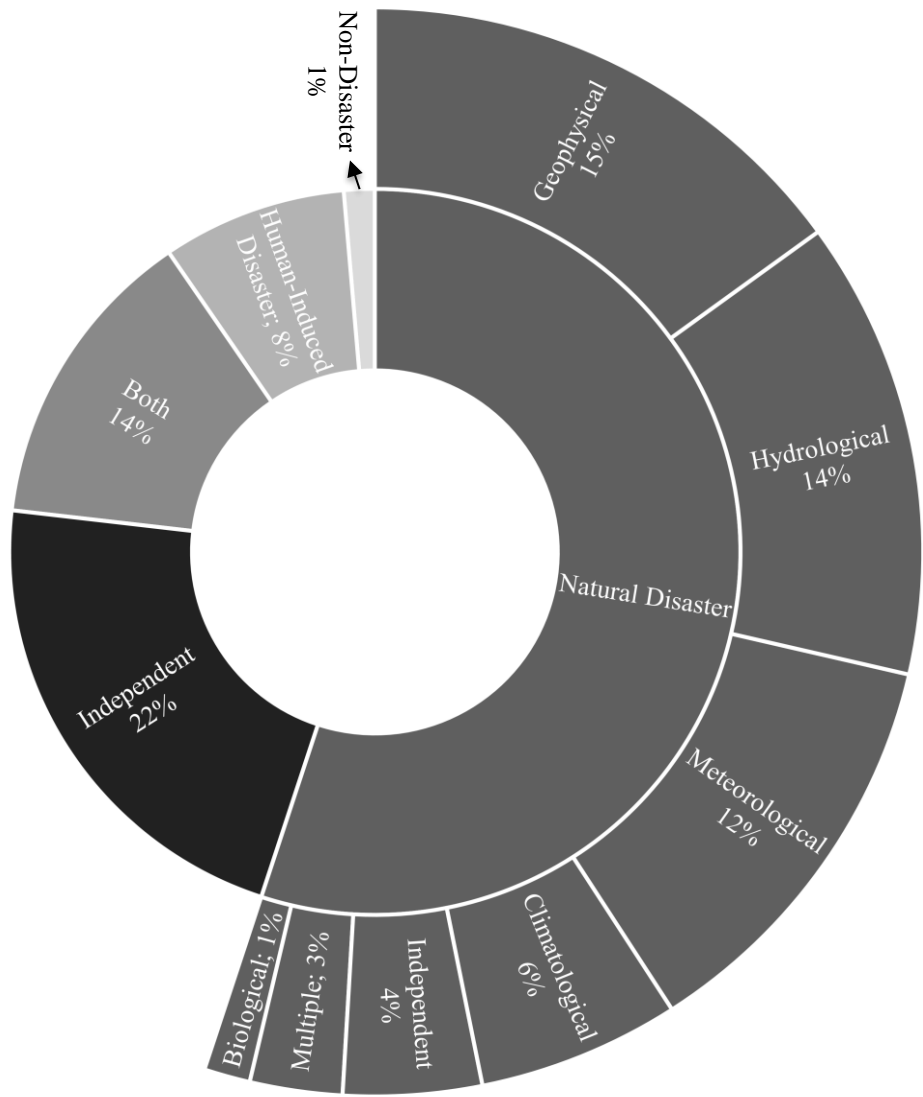

**Figure 3 Disaster cases in articles - separated by generic and first-level disaster group.
Source: Compilation by Author**

## 3.2. Disaster Phase

Disaster events in the extant research articles are divided into four phases; prevention, preparedness, response, and recovery (Cinnamon et al., 2016; Kankanamge et al., 2020b; Sarker et al., 2020). Figure 4 shows the distribution of articles between these four phases wherein the prevention or mitigation phase represents with bare 4%, following that preparedness stage which is slightly better than the initial phase but still less represented with 6%. Scholars' key priority over the years is on disaster response incidents, with 38% coverage. And the final phase recovery is less of interest to scholars with 1%, thereby demonstrating a drastic imbalance in research interest between the four stages. Additionally, 11% of the research focuses on more than one phase categorised as multiple. Remarkably, 29% of the research talks about the complete cycle of disaster with nearly half of them are conceptual papers. However, a sizeable portion of papers did not assess any single-phase as 11% of papers are located in the independent category. The independent category consists of articles that talk about the ethics of big data, big data in digital humanitarian practices, the hype around big data, and challenges, etc.

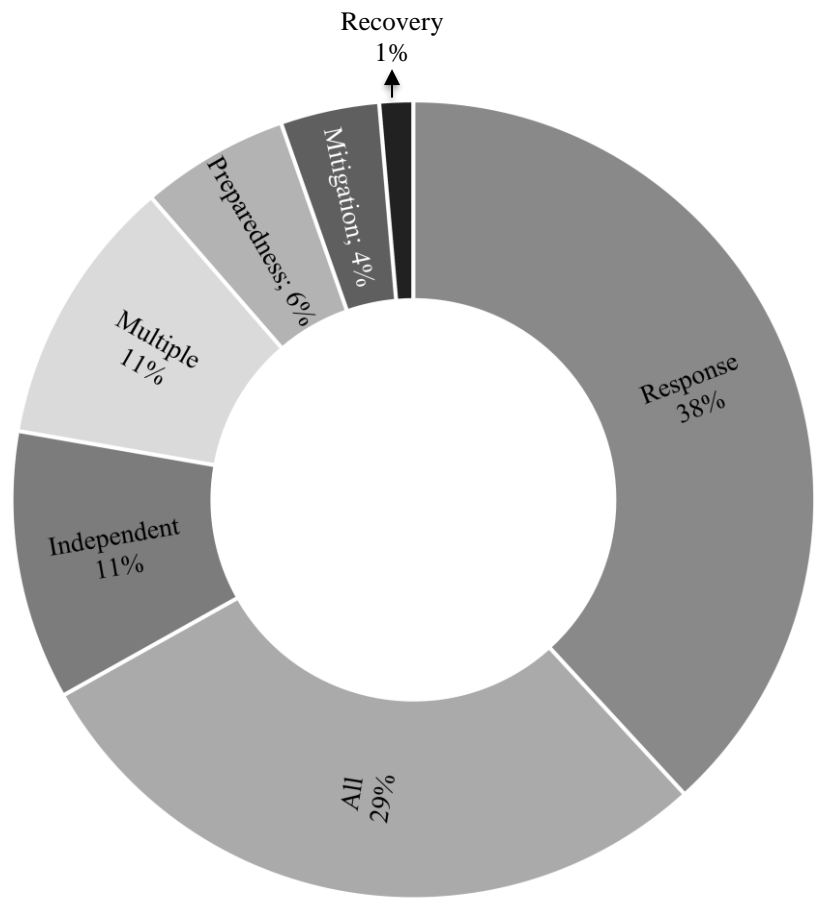

**Figure 4 Articles based on disaster phase. Source: Compilation by Author**

### 3.3. Disaster Locations

No region is immune to disasters, particularly to natural disasters, but some regions are seriously affected by both human casualties and economic losses. The Asian region still is vulnerable to disasters and ranks among the worst affected regions in the world (Swiss Re, 2019), and without any surprise, scholars favored examining the events in this region as shown in figure 5. The second most preferred region is the Americas but the United States is the country of choice for researchers turning blind eye to the South American region as 7 out of 8 papers focused on the United States with the main emphasis on hurricanes. The other regions Oceania, Europe, and Africa have received further less attention. Africa accounts for most human casualties to disasters second only to Asia (Swiss Re, 2019) but this region is least focused in this field where humanitarian and disaster assistance is paramount. Also, few studies are focused on multiple regions, but the large number is coming from the independent category which includes all the conceptual and some empirical papers where the research is not driven by the location.

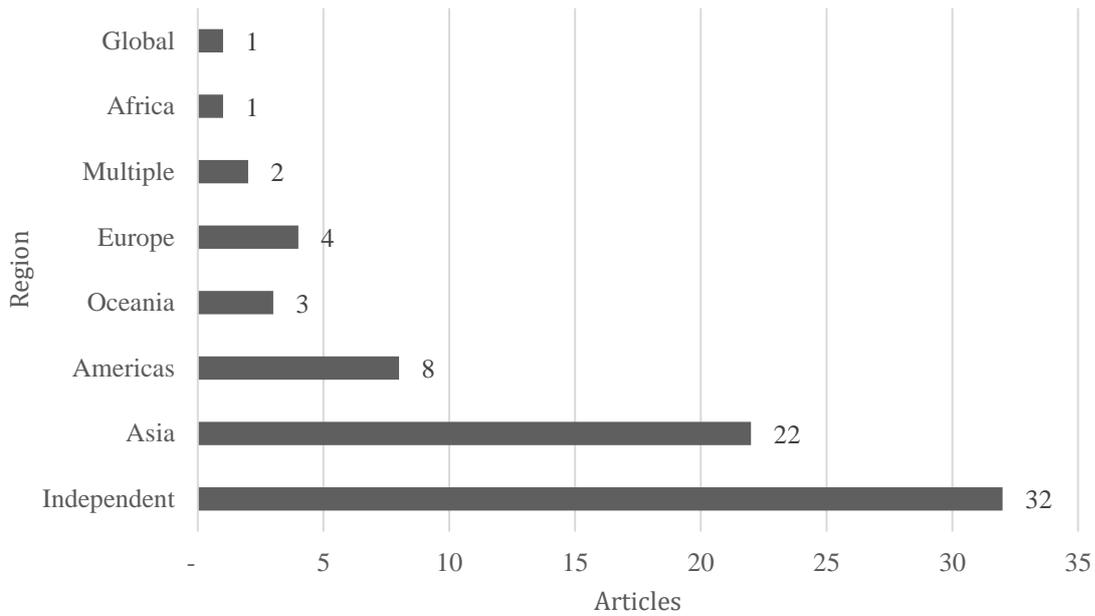

**Figure 5 Region of disaster in articles. Source: Compilation by Author**

Further, figure 5 displays the disasters by the year of occurrence. The disasters chosen by scholars are relatively new to the research timeframes, with an average time difference of 3.7 years between the disaster occurrence and the article publication, and these disasters occurred between 2011 and 2019. The year with the most scholarly papers (5) is 2012, which can be attributed to the gain of research interest in hurricanes in the United States. Similarly, a disaster that spans multiple years also represents most academic papers (5). The general category contains a huge number of articles as the studies with real disaster and humanitarian cases are minimal.

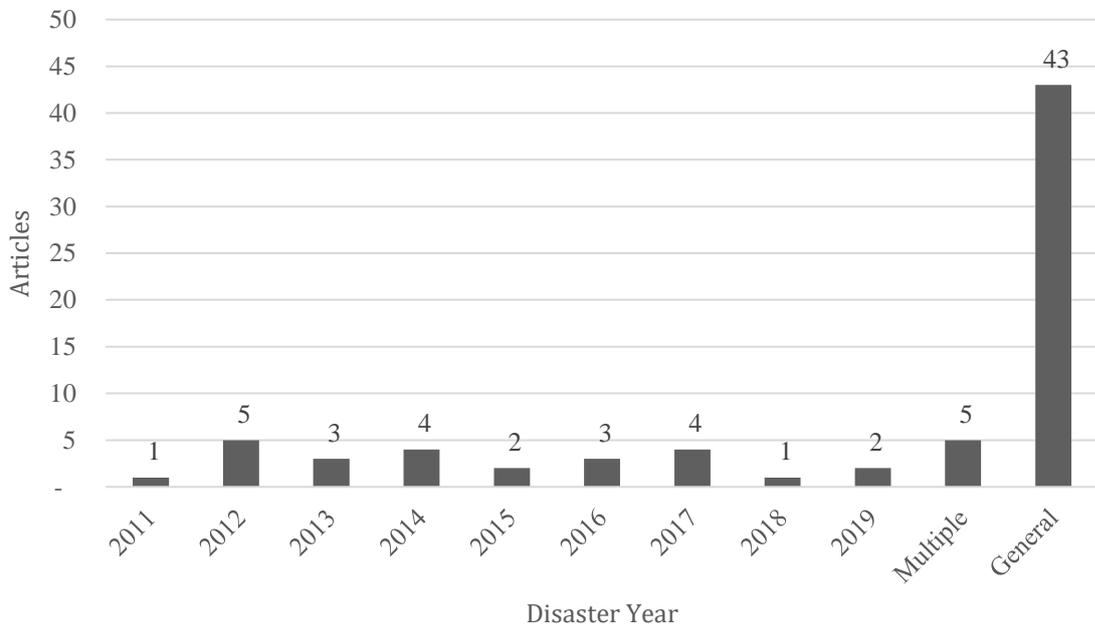

**Figure 6 Disaster years based on disaster cases in articles. Source: Compilation by Author**

### 3.4. Sources of Big Data

This study further assessed extant literature on the origins of data used for research, and figure 7 depicts all the sources of big data used. The least used data sources are authoritative data and crowdsourcing data with 1 article each. There are 2 studies each using maps data (such as micro maps and heat maps), spatial data, and satellite data. Mobile data, including positioning data and call detail records, is another source that is marginally better used, accounting for 4 articles. The most important disclosure is the use of social media data with a staggering 30 articles and this usage is not restricted to developed regions but utilized across all regions, covering all disaster phases and further applied in most of the disaster groups. Further, 10 articles studied multiple big data sources, and the remaining 21 are general articles without any focus on data sources.

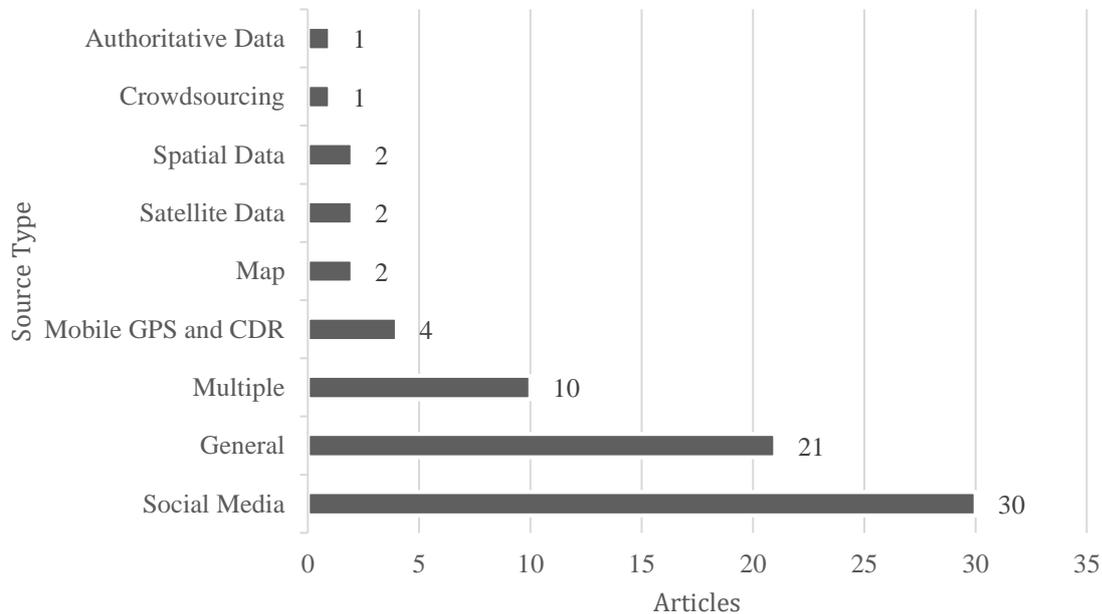

**Figure 7 Source of big data in articles. Source: Compilation by Author**

Within the social media data source, Twitter, and Weibo are preferred by the scholars in their research to find the solutions for HDO related challenges. The clear dominance of Twitter can be seen in figure 8 as it contains vast amounts of publicly accessible data that is easy to comprehend, and most significantly it offers timely data (Thom et al., 2016). However, over-dependent on Twitter might raise bias-related questions due to the heavy use of single social media (Avvenuti et al., 2018). Another social media Weibo presence is felt in the analysis with the representation of 4 articles but these studies are limited to the Asia region. Further, scholars used multiple social media data sources for their research in three instances and Twitter was the common source in all three studies. Authors claimed that the use of multiple social media platforms as a data source offers a holistic perspective of the disaster unfolding (Sherchan et al., 2017; Chaudhuri and Bose, 2020). Furthermore, one study conducted using Facebook as a data source.

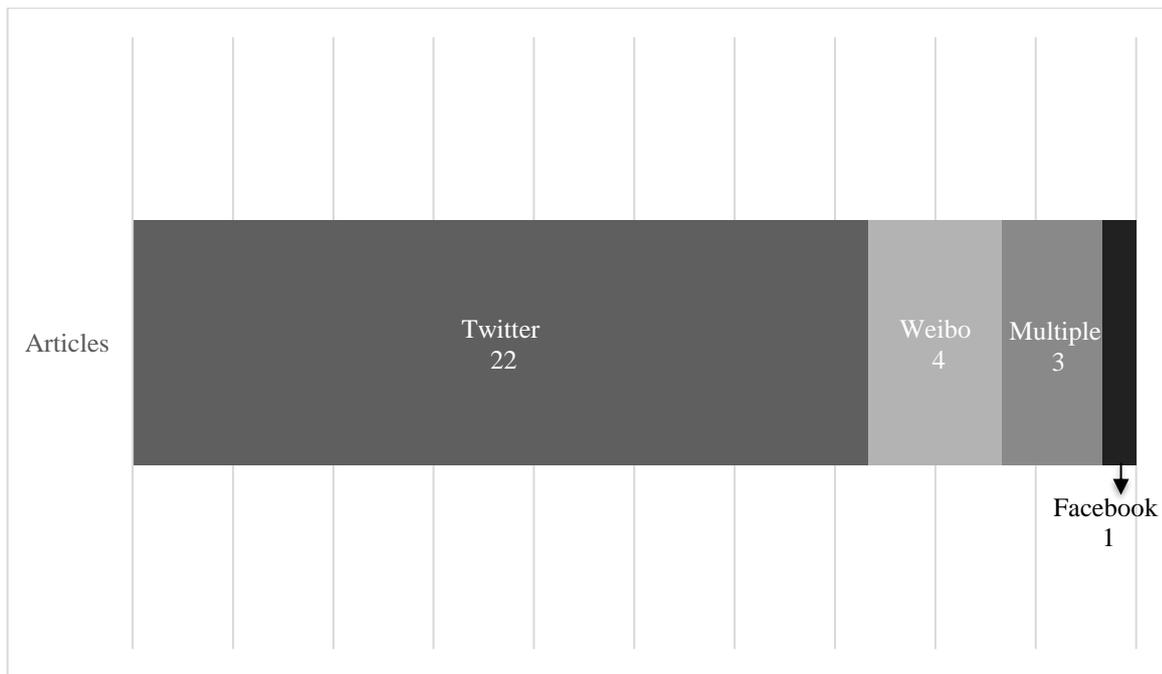

**Figure 8 Type of social media used in articles. Source: Compilation by Author**

### 3.5.   Theoretical Underpinnings

Theoretical foundations used in this field are listed in table 3. There is no clear domination in which theory is preferred, not only that no single theory has appeared more than once in the table, and a couple of papers used more than one theory. The minimal use of theory has been stated in Akter and Wamba's (2019) work but the improvement of theories inclusion can be seen in the last few years as the majority of theories mentioned in the below table are published in the last 3 years. The lesser number of theories in research papers may well be due to a much lower representation of the publications in the field of management.

| Theoretical Underpinnings in Articles | Author (Year) | No. of Articles |
|---|---|---|
| Local Model Semantics | Tomaszewski and MacEachren (2012) | n = 1 |
| Crowd Theory | Givoni (2016) | n = 1 |
| Information Theory | Read et al. (2016) | n = 1 |
| Organisational Mindfulness (OD) Theory, Information System Design Theory | Amaye et al. (2016) | n = 1 |
| Innovation Theory | Sandvik et al. (2017) | n = 1 |
| TOSE Resilience Framework (modified) | Papadopoulos et al. (2017) | n = 1 |
| Contingent Resource-Based View (CRBV) | Dubey et al. (2018) | n = 1 |
| Resource Dependence Theory | Prasad et al. (2018) | n = 1 |
| Social Exchange Theory | Li et al. (2018) | n = 1 |
| Disaster Damage Assessment Theory | Shan et al. (2019) | n = 1 |
| Organizational Information Processing Theory (OIPT), Swift Trust Theory | Dubey et al. (2019) | n = 1 |
| Resource-Based View Theory, Social Capital Theory | Jeble et al. (2019) | n = 1 |
| Social Support | Yan and Pedraza-Martinez (2019) | n = 1 |
| Technocolonialism | Madianou (2019) | n = 1 |
| Graph Theory | Warnier et al. (2020) | n = 1 |
| Organization Theory - Structural Dimensions | Fathi et al. (2020) | n = 1 |
| Theoretical Framework Developed: Critical Success Factor (CSF) approach | Susha (2020) | n = 1 |
| **Articles with explicit theoretical references** | | **n = 17** |
| **Articles without Theory** | | **n = 56** |
| **Total number of articles** | | **n = 73** |

**Table 3 Theories in Articles. Source: Compilation by Author**

## 4. DISCUSSION

The extant literature projects the dire need for more research into the field of BDA in HDO, this research highlighted areas that are crucial for the successful HDO. Information is and will be one of the deciding factors of success in relief operations mainly in logistics and supply chain (Warnier et al., 2020) as disaster response is the most intensive phase. However, there are few possible hurdles to cross: how reliable the information is, how quickly it is available to process for analysis, and how complete the data is. Time and again reliability is going to be a challenge, especially when using crowdsourcing and social media data because of false data or duplication in data (Kankanamge et al., 2020a; Nagendra et al., 2020; Rogstadius et al., 2013). The data is often needed to be available less than 72 hours after the disaster situation (van den Homberg et al., 2018). And, the disaster assessment is never going to be effective without the complete information (Shan et al., 2019). To begin with, the Humanitarian Data Exchange (HDX) may be a viable solution to two problems, data availability, and completeness. The OCHA platform has begun to create datasets for humanitarian and disaster response operations and now has 17,000 datasets with about 600,000 users (Centre for Humanitarian Data, 2020). However, this is a new initiative, and awareness within the humanitarian actors is essential so that the integrated datasets can be created for as many humanitarian and disaster situations as possible, thereby helping organizations make use of accurate data. Just 54% of these HDX datasets are complete as of now (Centre for Humanitarian Data, 2020). The completeness of these datasets and the continued addition of more locations would therefore be an excellent opportunity for relief operations and research.

Situational awareness, which helps in responding to sudden-onset disasters, is the other aspect that could contribute to a better response. Many emergency responders find that it is helpful to gain an understanding of the crisis in response (Zhang et al., 2020). All the sources of big data mentioned in figure 7, could improve situational awareness but social media data can often bring the ground reality, content dimension, and may well act as two-way communication between the affected people and relief coordinators (Wang and Ye, 2018; Zhang et al., 2019). But these sudden-onset disasters always put more pressure on disaster responders and at times this makes them opt for data that is less accurate because they don't want to lose the valuable time window.

Then there is the coordination factor, it is required whether the coordination between multiple organizations; within the country or international, or between multiple teams within the same organization. The coordination aspect is a big sell, Dubey et al.'s (2018) research discloses BDA can influence and further improves coordination in the humanitarian supply chain (HSC). Recently, more data is generating from crowdsourcing as well, meaning coordination is also required between organizations and individuals. This can be a challenge when digital humanitarians join hands with traditional humanitarian organisations due to the difference in working speed, hierarchical structures, and the variance in expertise (Sharma and Joshi, 2019).

Aside from the obvious emphasis on performance factors involved in the use of BDA in HDO, the other element surfaced in articles is social dimensions in a disaster setting. These are as important as any operational or managerial dimensions in HDO. Kontokosta and Malik's (2018) work on how big data can be helpful to reach the most affected people with a minimum capacity of resilience is noteworthy. Four domains in the social dimension; economic strength, social infrastructure, environmental conditions, and physical infrastructure are considered to measure neighborhood resilience. Authors developed an index called REDI which can rank the most resilient and least resilient areas. This will certainly help humanitarian and disaster responders to prioritise and reach the least resilient neighborhoods putting the relief supplies for effective use and reduce wastage in supplies and logistical efforts. Santos et al. (2020) combined multiple dimensions stating any one dimension is not sufficient enough for disaster response. Conceptual and a group of multiple dimensions workforce (W), economy (E), infrastructure (I), geography (G), hierarchy (H), and time (T) called WEIGHT includes a couple of social dimensions. Authors think that reliance through data analytics can only be improved with the integrated use of multiple dimensions. Also, social media is going to be an important source of information for HDO, not only because of large volumes of data but its timeliness which is crucial in disaster response and the two-way communication option makes it the most favourable choice. Using social media data, there is a possibility to analyse public emotions and sending psychological relief (Yang et al., 2019). But, there are some concerns in relying on social media data such as false information and lack of geo-tagging which helps in identifying the location is almost not available, basic 1% in the case of Twitter (Rogstadius et al., 2013). Also, location tags are not precise enough but these facts are becoming obsolete (Kankanamge et al., 2020b), and acting to strengthen these problems would be better than absolutely discarding social media for relief operations.

Primary findings indicate that scholars' interest in this field has risen to a 110% average change in research publications over the last decade which is derived from figure 2. The substantial growth has taken place over the last five years and 2015 can be seen as a turning point in this field. However, the major push is coming from computer science and engineering subject areas wherein management is far behind which needs to be balanced.

## 4.1. Future Research Directions

The avenues for future research are provided in table 4 from the viewpoint of big data sources where a single or different source of big data can be used to conduct the research.

| BD Source | Research Directions |
|---|---|
| Social Media | The location-routing problem (LRP) is one key area that can be hugely helpful for relief operations; either by delivering the supplies on time and potentially saving the unnecessary cost in logistics with effective route optimisation. Can big data, mainly social media be effective in developing LRP models? (Wu et al., 2020) |
| Social Media | The time dimension is the least explored area out of four dimensions of social media data, the other three; network, content, and space dimensions are comparatively well researched. Further, it would be worthwhile to consider all four dimensions simultaneously while using social media as a source of big data (Wang and Ye, 2018). |
| Social Media | The behaviour and sentiment of the affected individuals are areas less studied. A crisis data set that can be used as a standard guide for disasters is also not available. The creation of this data set would help identify the needs of the affected people by using social media data such as Twitter and help to create a sentiment model (Ragini et al., 2018). |
| Social Media | What are the causal effects between social media attention and social characteristics such as income, education, and wealth in disaster-affected areas? Would the use of substantial BDA in analysing the social media data to deliver the relief efforts will be biased? (Fan et al., 2020) |
| Crowdsourcing | How will geographically dispersed crowdsourcing networks be impacted by the diverse geopolitics involved in humanitarian assistance? Can digital humanitarians change the direction of this interplay and affect geopolitics with the addition of BDA? (Mulder et al., 2016) |
| Different Sources | The ever-increasing volume of big data offers possibilities that we never thought of two decades ago. Lin et al.'s (2020) work shows that we can now estimate the demand for relief supplies with the use of big data in the disaster response stage, this research has been carried out using Baidu. This will certainly help to save the relief wastage and logistical expenditure. Future work can be conducted using other sources of big data in various disasters and compare their effectiveness. |
| Different Sources | How will the efficacy of the HSC be impacted by Big Data Predictive Analytics along with social capital? (Jeble et al., 2019) |
| Different Sources | Considering BDA as organisational capability, research improvements in examining the intangible resource such as organisational culture have begun (Dubey et al., 2019). This needs to be expanded further to another intangible resource organisational learning and how the culture change will affect the use of BDA in HDO. |
| Different Sources | How do humanitarian agencies assign priority to the population at risk in delivering the aid? Can this be enhanced by incorporating data analytics? |

**Table 4 Future directions to research in the field of BDA and HDO. Source: Compilation by Author**

The most hard-hitting disaster groups in recent years are climatological and biological, but they have been relatively uninteresting to scholars with a mere representation in the reviewed papers. In addition to the research perspectives listed in the above table, scholars must concentrate on these less explored disaster groups within the natural disaster generic group along with the human-induced disaster generic group to see how BDA can be beneficial. Further increase in the research gap between both disaster generic groups could lead to uneven arguments and justifications of BDA in the HDO spectrum. The disparity of research in disaster locations is much wider and concerning as developed economies got more attention than Africa, the poorest region in the global south. Africa accounts for most human casualties to disasters second only to Asia (Swiss Re, 2019) but this region is least focused in this field where humanitarian and disaster assistance is paramount.

## 5.     CONCLUSION

At the beginning of this study, 168 million people required humanitarian assistance (beginning of 2020), by the end of completing this review the number rose to 235 million (at the end of 2020). There is no time to lose, certainly no data to lose. Humanitarian organizations must take advantage of BDA with the same earnestness as profit-oriented organisations by keeping the checks on ethical concerns. Providentially, academic research is expanding quickly in this spectrum especially in the last 2 years of research contributing more than 50% of overall research. There is also similar growth in the management subject area but the overall contribution to the field of BDA in HDO is trivial. As it is a multidisciplinary field, the contribution of other subject areas is often significant but the management subject area needs to catch up and increase the presence. This review is attempted to approach the topic and three research questions in a more integrated and systematic way. First, over the past couple of years, research on the use of BDA for HDO has significantly improved, showing the ability of researchers to explore what data analytics can do to enhance the way humanitarian and disaster relief operates. Management research, though, is well behind and fragmented in its contribution to the field. Second, the status of BDA application across different disaster categories, disaster phases, and disaster locations is imbalanced and research priorities are not utilised where it is more essential. In disaster phases, concentrating only on disaster response while overlooking the other three stages; prevention, mitigation, and recovery would not result in a holistic improvement of the field. Furthermore, there is a high dependency on social media as a source of big data, which poses ethical, bias, and factual concerns that must be addressed. Third, the dearth of theoretical frameworks is evident in the field, while this seems to be improving lately, the proportion of papers with a theoretical lens in overall papers in each year is not encouraging. Irrespective of these critical findings, the review does have some limitations and the author is aware of these limitations during stage one in step three while designing the review protocol.

## 5.1. Limitations

There are two major limitations, one in database selection and the second in exclusion criteria which is not part of five-level search criteria. Although, database selection is logical in this review if the time and resources are allowed web of science can also be included for future reviews, which might bring a few more papers to the review process and provide a much clearer view of the topic. The second limitation is more of a feature related to the Scopus. The database has two options to refine the results namely 'Limit to' and 'Exclude'. In the subject area, one of the refine results options in Scopus, won't provide a unique breakdown for articles using 'Limit to' criteria. The reason is Scopus assigns each article to multiple subject areas making it impossible to get the unique numbers for each subject area when using the 'Limit to' criteria. Author rationale for using this criterion rather than 'Exclude to' is because the 'Limit to' feature eliminates all the documents which have subject areas mentioned in the 'Exclude to' list, including subject areas that the author is interested in but are excluded because of multiple subject area tags to each document.